\documentclass{ws-ijmpd}
\usepackage[super,compress]{cite}
\usepackage{amssymb}
\usepackage{pxfonts}

\begin{document}
\markboth{Vyacheslav Dokuchaev}
{To see the invisible: image of the event horizon within the black hole shadow}

%
\catchline{}{}{}{}{}
%

\title{TO SEE THE INVISIBLE: IMAGE OF THE EVENT HORIZON \\ WITHIN THE BLACK HOLE SHADOW}
\author{VYACHESLAV DOKUCHAEV}
\address{Institute for Nuclear Research of the Russian Academy of Sciences \\
	Prospekt 60-letiya Oktyabrya 7a, Moscow, 117312, Rissia \\ \vskip0.2cm
	National Research Nuclear University ``MEPhI'' (Moscow Engineering Physics Institute) \\
	Kashirskoe shosse 31, Moscow, 115409, Russia  \\
	dokuchaev@inr.ac.ru}
\maketitle
\begin{history}
\received{13 December 2018}
\revised{Day Month Year}
\end{history}

\begin{abstract}
How	the supermassive black hole SgrA* in the Milky Way Center looks like for a distant observer? It depends on the black hole highlighting by the surrounding hot matter. The black hole shadow (the photon capture cross-section) would be viewed if there is a stationary luminous background. The black hole event horizon is invisible directly ({\it per se}). Nevertheless, a more compact (with respect to black hole shadow) projection of the black hole event horizon on the celestial sphere may be reconstructed by detecting the highly red-shifted photons emitted by the non-stationary luminous matter plunging into the black hole and approaching to the event horizon. It is appropriate to call this reconstructed projection of the event horizon on the celestial sphere for a distant observer as  the ``{\sl lensed event horizon image}'', or simply the ``{\sl event horizon image}''. This event horizon image is placed on the celestial sphere within the position of black hole shadow. Amazingly, the event horizon image is a gravitationally lensed projection on the celestial sphere of the whole surface of the event horizon globe. In result, the black holes may be viewed at once from both the front and back sides. The lensed event horizon image may be considered as a genuine silhouette of the black hole. For example, a dark northern hemisphere of the event horizon image is the simplest model for a black hole silhouette in the presence of a thin accretion disk.
\end{abstract}
\ccode{PACS numbers: 04.70.Bw, 98.35.Jk, 98.62.Js.}

\tableofcontents

\section{Introduction}

Nowadays the LIGO detectors present the crucial evidence for the existence of black holes in the Universe by direct observations of gravitational waves from merging of massive black holes in binary systems \cite{LIGO16a,LIGO16b,LIGO16c,LIGO16d,LIGO16e}. At the same time these direct observations of gravitational waves provide the first verification of General Relativity in the strong field limit. The other possibility for a detailed  verification of General Relativity in the strong field limit is in observations of the nearest environment of supermassive black holes.

The supermassive black hole SgrA* at the Galactic Center with a mass $M=(4.3\pm0.3)10^6M_\odot$ is under the intensive investigations by different instruments and methods as from the Earth surface and also from the space \cite{Ghez08,Gillessen09a,Gillessen09b,Meyer12,Johnson15,Chatzopoulos15,Eckart17,Parsa17,Johnson18}. One reason is that this supermassive black hole is the nearest ``dormant'' quasar dwelling at the center of our native Galaxy. The other reason is that modern scientific technologies permit to view the nearest environment of the event horizon of this enigmatic black hole directly for the first time. A challenging ``The Event Horizon Telescope Project'', which is a global network of mm and sub-mm telescopes \cite{Fish16,Lacroix13,Kamruddin,Johannsen16,Johannsen16b,Broderick16,Chael16,Kim16,Roelofs17a,Roelofs17b,Doeleman17}, intends to reveal the silhouette of SgrA* \cite{Synge,49,Chandra,51,52,Falcke13,57,Cunha15,Abdujabbarov15,Johannsen16c,Younsi16,70,Thorne2015,doknaz18d}. If successful, the result of this project would be the first direct experimental verification (or falsification) of the very existence of black holes in the Universe. 

A black hole shadow is the surface of photon capture cross-section viewed by distant observers on the celestial sphere due to the gravitational lensing of luminous background, consisting of bright stars or hot gas, located either far enough from the black hole or at the stationary orbits around it (for details see, e.\,g., Refs.~\refcite{Grenzebach14,Grenzebach15,Cunha18a,Cunha18b,Huang18} and references therein). 

A next scientific goal in investigation of the black hole SgrA* would be the detailed elaboration of its silhouette for verification not only the General Relativity in the strong field limit but also its numerous modifications  \cite{Grenzebach14,Grenzebach15,Cunha18a,Cunha18b,Huang18,71,65,Dokuch14,DokEr15,FizLab,Nucamendi15,Nucamendi16,CliffWill17a,CliffWill17b,Goddi17,Mureika17,Zakharov18a,Zakharov18b,Zakharov18c,Lamy18,Zajacek18}. The promising first step for this future study is the Millimetron space interferometer project with an angular resolution of about one nanoarcsecond \cite{64}. 

In this paper we describe the possibility for indirect observation of the  black hole event horizon. The black hole event horizon is invisible directly ({\it per se}). Nevertheless, we demonstrate below that gravitational lensing of the luminous matter plunging into the black hole provides the principle possibility for reconstruction of the black hole event horizon projection on the celestial sphere for a distant observer. It is appropriate to call this reconstructed projection of the event horizon on the celestial sphere for a distant observer as  the ``{\sl lensed event horizon image}'', or simply the ``{\sl event horizon image}''. 

\section{Specific features of geodesics in Kerr metric}

The standard form of Kerr metric in the Boyer-Lindquist coordinate system \cite{BoyerLindquist} with coordinates $(t,r,\theta,\phi)$
is
\begin{equation}
ds^2=\frac{\Sigma\Delta}{\mathcal A}dt^2
-\frac{{\mathcal A}\sin^2\theta}{\Sigma}(d\phi-\omega dt)^2-
\frac{\Sigma}{\Delta}dr^2-\Sigma d\theta^2,
\label{metric}
\end{equation}
where
\begin{eqnarray}
\Delta &= & r^2-2Mr+a^2, \label{Delta} \\
\Sigma &=& r^2+a^2\cos^2\theta,  \label{Sigma} \\
{\mathcal A} &=& (r^2+a^2)^2-a^2\Delta\sin^2\theta, \label{A} \\
\omega &=& \frac{2Mr}{\mathcal A}\, a.
\label{omega}
\end{eqnarray}
In these equations $M$ is a black hole mass, $a$ is a specific angular momentum of the black hole and $\omega$ is the angular dragging velocity. It is used units: the gravitational constant $G=1$ and the velocity of light $c=1$.

The event horizon of the Kerr black hole, $r=r_{\rm h}$, is the larger root of the equation $\Delta=0$, i.\,e., 
\begin{equation}
r_{\rm h}=M+\sqrt{M^2-a^2}.
\label{rh}
\end{equation}
The event horizon exists only if $M^2\geq a^2$. At $M^2<a^2$ the metric (\ref{metric}) describes a naked singularity. The black hole event horizon rotates rigidly as a solid body (i.\,e., independent of the latitude coordinate $\theta$) with an angular velocity \cite{smarr,bch}
\begin{equation}
\Omega_{\rm h}\equiv\lim_{r\rightarrow r_{\rm h}}\omega 
=\frac{a}{2M r_{\rm h}},
\label{Omegah}
\end{equation}
where a dragging angular velocity in Kerr metric $\omega$ is from Eq.~(\ref{omega}).

The motion of test particles with a mass $\mu$ in Kerr metric is completely defined by three integrals of motion \cite{Carter68,deFelice}: the total particle energy $E$, the azimuthal component of the angular momentum $L$ and the Carter constant $Q$, which is related with a total angular momentum of the particle. The Carter constant is zero, $Q=0$ for trajectories, which confined in the black hole equatorial plane. The total angular momentum of the particle is $\sqrt{Q+L^2}$ in the particular case of nonrotating black hole with $a=0$.

The trajectories of test particles are governed by equations of motion \cite{Chandra,Carter68,deFelice,BPT,MTW}:
\begin{eqnarray}
\Sigma\frac{dr}{d\tau} &=& \pm \sqrt{V_r}, \label{rmot} \\
\Sigma\frac{d\theta}{d\tau} &=& \pm\sqrt{V_\theta}, \label{thetamot} \\
\Sigma\frac{d\varphi}{d\tau} &=& L\sin^{-2}\theta+a(\Delta^{-1}P-E),
\label{phimot} \\
\Sigma\frac{dt}{d\tau} &=& a(L-aE\sin^{2}\theta)+(r^2+a^2)\Delta^{-1}P,
\end{eqnarray}
where 
\begin{eqnarray}
V_r &=& P^2-\Delta[\mu^2r^2+(L-aE)^2+Q], \label{Vr} \\
V_\theta &=& Q-\cos^2\theta[a^2(\mu^2-E^2)+L^2\sin^{-2}\theta],
\label{Vtheta} \\
P &=& E(r^2+a^2)-a L, \\
\Sigma &=& r^2+a^2\cos^2\theta, \\\
\Delta &=& r^2-2r+a^2.  \label{Delta2}
\end{eqnarray}
In these equations $\lambda$ is an affine parameter. The proper time along a timelike geodesic is $\mu\tau$ and a null geodesic has $\mu=0$.

The effective radial $V_r$ and latitudinal $V_\theta$ potentials in (\ref{Vr}) and (\ref{Vtheta}) define the motion of particles in $r$- and $\theta$-directions. 

For numerical calculations of gravitational lensing in the Kerr space-time background we use also the integral equations of motion for photons \cite{Chandra,Carter68,deFelice,BPT,MTW}: 
\begin{equation}\label{eq2425}
\fint^r\frac{dr}{\sqrt{V_r}}
=\fint^\theta\frac{d\theta}{\sqrt{V_\theta}},
\end{equation}
\begin{equation}
\phi=\fint^r\frac{aP}{\Delta\sqrt{V_r}}\,dr
+\fint^\theta\frac{L-aE\sin^2\theta}{\sin^2\theta\sqrt{V_\theta}}\,d\theta, \label{eq25bbb} 
\end{equation}
\begin{equation}
t=\fint^r\frac{(r^2+a^2)P}{\Delta\sqrt{V_r}}\,dr
+\fint^\theta\frac{(L-aE\sin^2\theta)a}{\sqrt{V_\theta}}\,d\theta, \label{eq25tt} 
\end{equation}
\begin{equation}\label{eq2425b}
\tau=\fint^r\frac{r^2}{\sqrt{V_r}}\,dr
+\fint^\theta\frac{a^2\cos^2\theta}{\sqrt{V_\theta}}\,d\theta,
\end{equation}
The integrals in (\ref{eq2425})--(\ref{eq25tt}) are the path integrals along a trajectory connecting the luminous source and distant observer.

We will use in the following mainly the normalized dimensionless variables and parameters: $r\Rightarrow r/M$, $a\Rightarrow a/M$, $e\Rightarrow e/M$,  $\epsilon\Rightarrow \epsilon/\mu$, $E\Rightarrow
E/\mu$, $L\Rightarrow L/(M\mu)$, $Q\Rightarrow Q/(M^2\mu^2)$. 

With these normalized dimensionless variables and parameters the corresponding trajectories of test massive particles (with a mass $\mu\neq0$) in the Kerr space-time are determined by the three integrals of motion: $\gamma=E/\mu$, $\lambda=L/E$ and $q=\sqrt{Q}/E$. 

The trajectories of photons (null geodesics) in Kerr metric are determined by only two parameters, $\lambda$ and $q$, which are related with the horizontal $\alpha$ and vertical $\beta$ impact parameters of photons on the celestial sphere viewed by a distant observer (see definitions below).

\begin{figure}[pt]
	\centerline{\psfig{file=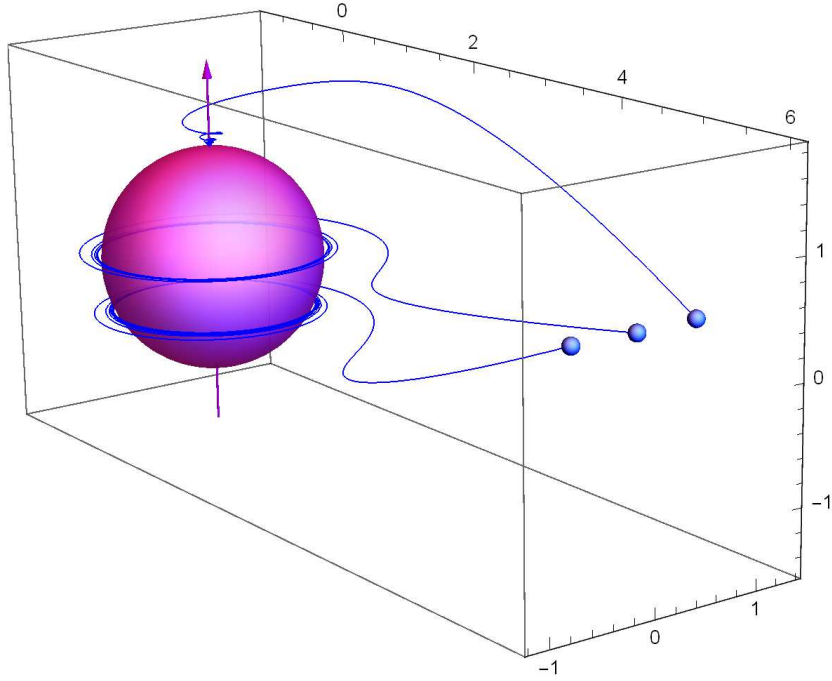,width=11cm}}
	\vspace*{8pt}
	\caption{$3D$ trajectories of massive test particles ($\mu\neq0$), plunging into the rotating Kerr back hole with $a=0.998$. The first particles (with $\gamma=1$, $\lambda=0$, $q=1.85$) is plunging near the north pole of the event horizon. The second particle (with $\gamma=1$, $\lambda=-1.31$, $q=0.13$) is plunging near the equator. At last, the third one (with $\gamma=1$, $\lambda=-1.31$, $q=0.97$) is plunging in the south hemisphere. \label{f1}}
\end{figure}

See in Refs.~\refcite{Gralla15,Strom16,Gralla16,Strom17,Strom18} some examples of analytical computations for null geodesics in the Kerr space-time. 


In this paper it is used the classical Boyer--Lindquist coordinate system \cite{BoyerLindquist} with coordinates $(t,r,\theta,\phi)$. Although in the Boyer-Lindquist coordinates all the surfaces $r=const$ are confocal ellipsoids \cite{BoyerLindquist,Chandra}, for transparency of pictures we plot the event horizon  $r_{\rm H}=const$ as a sphere  in our 3D Figs. 1--4 and 6 using the fact that topologically it is the 2-sphere \cite{BoyerLindquist,Chandra}. The local Gaussian curvature of the event horizon in Kerr metric depends on the latitude coordinate $\theta$. In particular, the form of black hole horizon embedded in our familiar Euclidean 3D-space is an oblate ellipsoid of revolution in the case of a slow rotation ($a<<1$). Surprisingly, the event horizon of a fast rotating black hole ($2\sqrt{2}/3<a\leq1$) has the negative Gaussian curvature both on and around the axis of symmetry, and the event horizon of a fast rotating black hole cannot be globally embedded in Euclidean 3D-space at all \cite{Smarr73,Sharp81}. 


It must be especially stressed that forms of both the black hole shadow and the event horizon image, viewed by a distant observer (at Newtonian limit, $r\gg r_{\rm H}$), are independent on the used coordinate system (see 2D Figs. 5, 7--10).

In Fig.~1 are shown some representative examples of $3D$ trajectories for massive test particles, plunging into rotating back hole and infinitely ``winding'' around black hole with an angular velocity (\ref{Omegah}) by approaching to the event horizon at $r=r_{\rm h}$. These three trajectories are reproduced from combined numerical solution of integral equations of motion (\ref{eq2425}) and (\ref{eq25bbb}).

\section{Black hole shadow on the stationary luminous background}

\begin{figure}[pt]
\centerline{\psfig{file=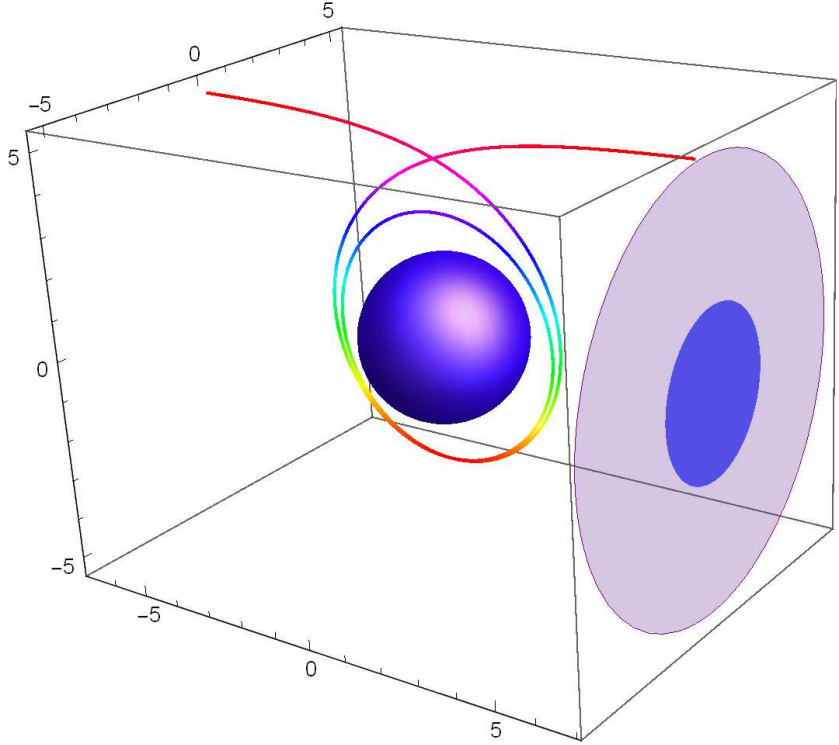,width=11cm}}
\vspace*{8pt}
\caption{The black hole shadow projection on the celestial sphere (purple disk with a radius $r_{\rm sh}=3\sqrt{3}\simeq5.196$) in the Schwarzschild case, $a=0$. Inside the shadow there it is pictured a fictitious image (blue disk with a radius $r_{\rm h}=2$) of the black hole event horizon in the fictitious Euclidean space. It is shown the representative photon trajectory (multicolored $3D$ curve) with impact parameters $\lambda=0$ and $q=3\sqrt{3}$. This photon is starting from a distant background, then winding near a radial turning point at $r_{\rm min}=3$ around the black hole event horizon (blue sphere), and finally is reaching a distant observer at the north pole of the black hole shadow projection on the celestial sphere. \label{f2}}
\end{figure}

\begin{figure}[pt]
\centerline{\psfig{file=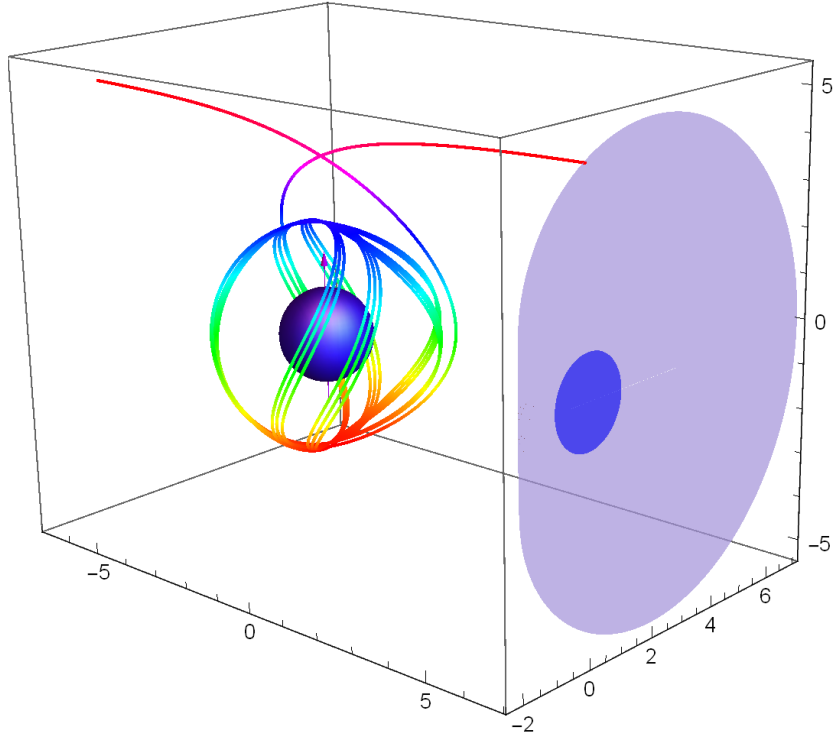,width=11cm}}
\vspace*{8pt}
\caption{The black shadow projection on the celestial sphere (purple region) in the extreme Kerr black hole case, $a=1$. It is shown also a representative example of the photon trajectory (multicolored $3D$ curve) near the shadow boundary (with $\lambda=0$ and  $q=\sqrt{3-\sqrt{2}}(1+\sqrt{2})^{3/2}]\simeq4.72$). This photon is starting from a distant background, then winding around the black hole event horizon (blue sphere), and, finally, is reaching a distant observer at the intersection of the shadow boundary with the black hole rotation axis projection on the celestial sphere. The corresponding radial turning point of this photon is at $r_{\rm min}=1+\sqrt{2}$. A black hole rotation axis is shown by vertical magenta arrow crossing the event horizon sphere. Inside the shadow it is  pictured a fictitious image (blue disk with a radius $r_{\rm h}=1$) of the black hole event horizon in the fictitious Euclidean space. \label{f3}}
\end{figure}

A black hole shadow in Kerr metric, projected on the celestial sphere and viewed by a distant observer in the equatorial plane of the black hole, is determined from simultaneous solution of equations $V_r(r)=0$ and $[rV_r(r)]'=0$, where the effective radial potential $V_r(r)$ is from Eq.~(\ref{Vr}). The corresponding solution for black hole shadow in the parametrical form $(\lambda,q)=(\lambda(r),q(r))$ is
\begin{eqnarray} \label{shadow1}
\lambda&=&\frac{-r^3+3r^2-a^2(r+1)}{a(r-1)}, \\
q^2&=&\frac{r^3[4a^2-r(r-3)^2]}{a^2(r-1)^2}.
\label{shadow2}
\end{eqnarray}
(see, e.\,g., Refs.~\refcite{49,Chandra} for more details).

The black hole shadow projection on the celestial sphere is shown in Fig.~2 for the Schwarzschild case, $a=0$. The radius of this shadow (purple disk) is $r_{\rm sh}=3\sqrt{3}\simeq5.196$. The corresponding shadow of the extreme Kerr black hole $(a=1)$ is shown in Fig.~3. 

Positions of the gravitationally lensed photons on the celestial sphere are determined by two impact parameters, $\alpha$ --- the apparent displacement of the image perpendicular to the projected black hole rotation axis and $\beta$ --- the apparent displacement parallel to the projected black hole rotation axis \cite{CunnBardeen73}:
\begin{eqnarray} 
\alpha&=&=-\frac{\lambda}{\sin\theta_0}, 
\\
\beta&=&q^2+a^2\cos^2\theta_0-\lambda^2\cot^2\theta_0,
\end{eqnarray}
where $\cos\theta_0$ is a latitude coordinate of the distant observer. In numerical calculation we choose $\cos\theta_0=0.1$, $\theta_0\simeq84^\circ\!\!.\,24$.

\section{Event horizon image elicited by the plunging luminous sources}

Let us imagine that numerous luminous probes plunge into the black hole from different directions and approach to the black hole event horizon with a permanently growing redshift. The hot compact gas clouds and/or neutron stars would be the best candidates for the described plunging luminous probes in the real astrophysical conditions near the supermassive black hole SgrA*. The gravitationally lensed images of these probes, approaching to the event horizon, may be detected by a distant observer. The last hopefully detected lensed photon, emitted by a separate plunging probe in the vicinity of the event horizon, reconstructs on the celestial sphere the position of  specific point in the vicinity of the event horizon. By using this procedure for a set of different plunging probes, it would be possible to reconstruct the whole projection of the invisible event horizon on the celestial sphere. The accuracy of this reconstruction will depend on the sensitivity of detectors for highly red-shifted photons, emitted by the plunging probes near the event horizon. In result, it would be obtained the lensed image of the event horizon on the celestial sphere.

\begin{figure}[pt]
\centerline{\psfig{file=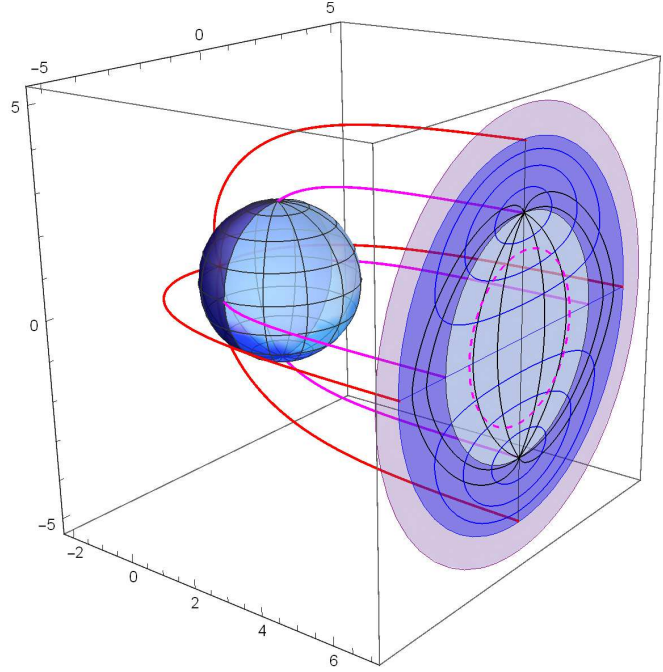,width=11cm}}
\vspace*{8pt}
\caption{The lensed event horizon image on the celestial sphere (dark-blue and light-blue disks) within the black hole shadow (purple disk) in the Schwarzschild case, $a=0$. The characteristic trajectories of photons forming the event horizon image on the celestial sphere are shown: four (red colored) trajectories of photons with $\sqrt{\lambda^2+q^2}=r_{\rm eh}$ are starting from the farthest point of the event horizon globe, and, respectively, four (magenta colored) trajectories of photons with $\sqrt{\lambda^2+q^2}=r_{\rm EW}$ are starting from the East-West meridian. Images of parallels (blue curves) and meridians (black curves) are shown on the event horizon globe (blue sphere) and on their corresponding projected images on the celestial sphere (dark-blue and light-blue regions). The nearest part (to a distant observer) of the event horizon hemisphere is shown by light-blue color. Respectively, the farthest part of the event horizon hemisphere is shown by the dark-blue color. \label{f4}}
\end{figure}

\begin{figure}[pt]
\centerline{\psfig{file=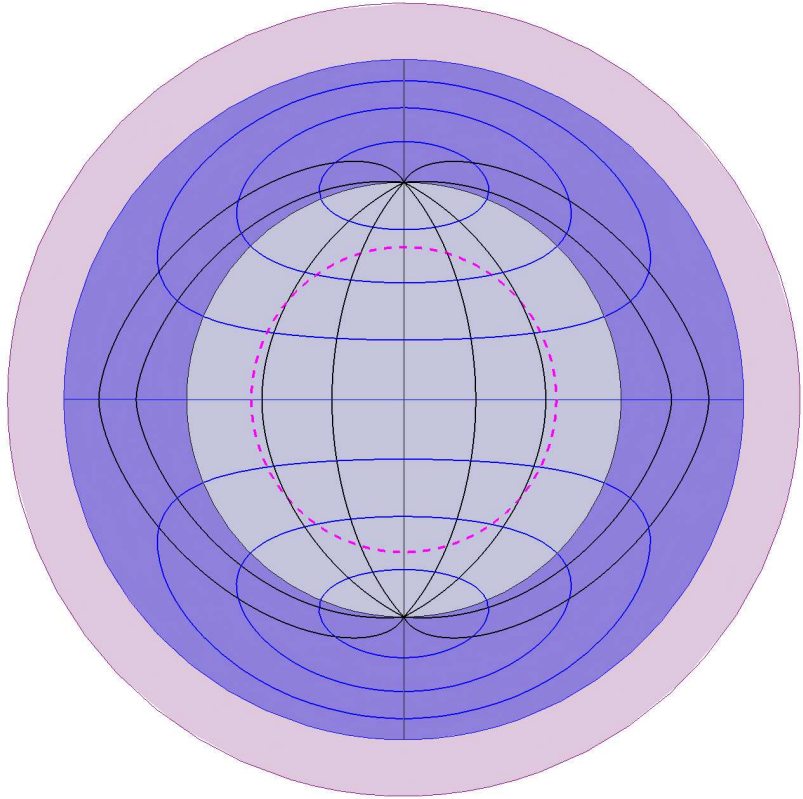,width=11cm}}
\vspace*{8pt}
\caption{The lensed event horizon image projected on the celestial sphere in the Schwarzschild case, $a=0$ (see detailed description in Fig.~\ref{f4}. \label{f5}}
\end{figure}

\begin{figure}[pb]
\centerline{\psfig{file=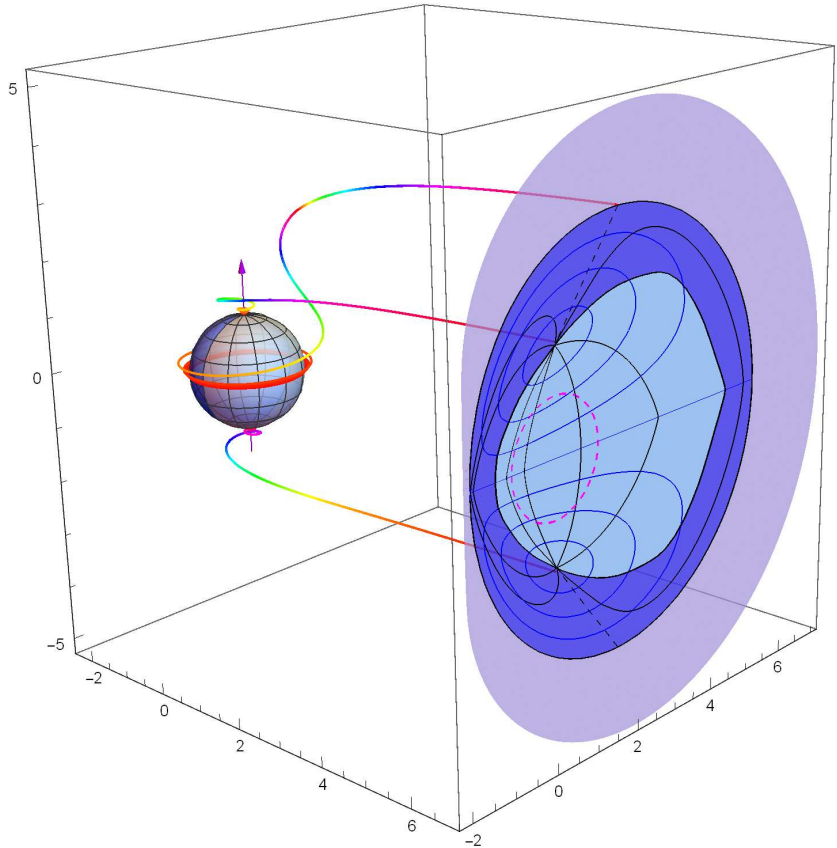,width=11cm}}
\vspace*{8pt}
\caption{The lensed event horizon image on the celestial sphere (dark-blue and light-blue disks) within the black hole shadow (purple disk) in the extreme Kerr black hole case, $a=1$. The characteristic trajectories of photons  (multicolored $3D$ curves) forming the event horizon image on the celestial sphere are shown: the photons outgoing to a distant observer from the North and South poles (with $\lambda=0$, $q = 1.77$), and, additionally, the photon starting from the equator (with $\lambda=-1.493$, $q=3.629$) of the event horizon globe with a radius $r_{\rm h}=1$. The black hole shadow is the largest purple region on the celestial sphere. Images of parallels (blue curves) and meridians (black curves) are shown on the event horizon globe (blue sphere) and on their corresponding projected images on the celestial sphere (dark-blue and light-blue regions). The nearest part (to a distant observer) of the event horizon hemisphere is shown by light-blue color. Respectively, the farthest part of the event horizon hemisphere is shown by the dark-blue color. The dashed curve is an instant prime meridian. \label{f6}}
\end{figure}

\begin{figure}[pt]
\centerline{\psfig{file=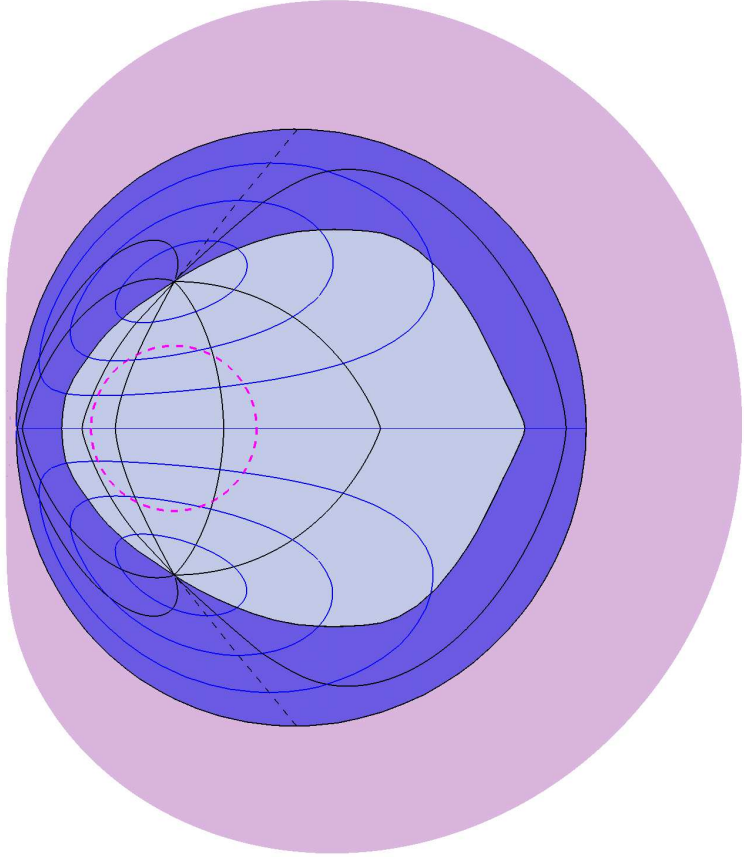,width=11cm}}
\vspace*{8pt}
\caption{The projected event horizon image on the celestial sphere in the extreme Kerr case, $a=1$. As well as in the Schwarzschild case a distant observer views the reconstructed event horizon image at once from both the front and back sides. (see detailed description in Fig.~\ref{f6}). \label{f7}}
\end{figure}

\begin{figure}[pt]
\centerline{\psfig{file=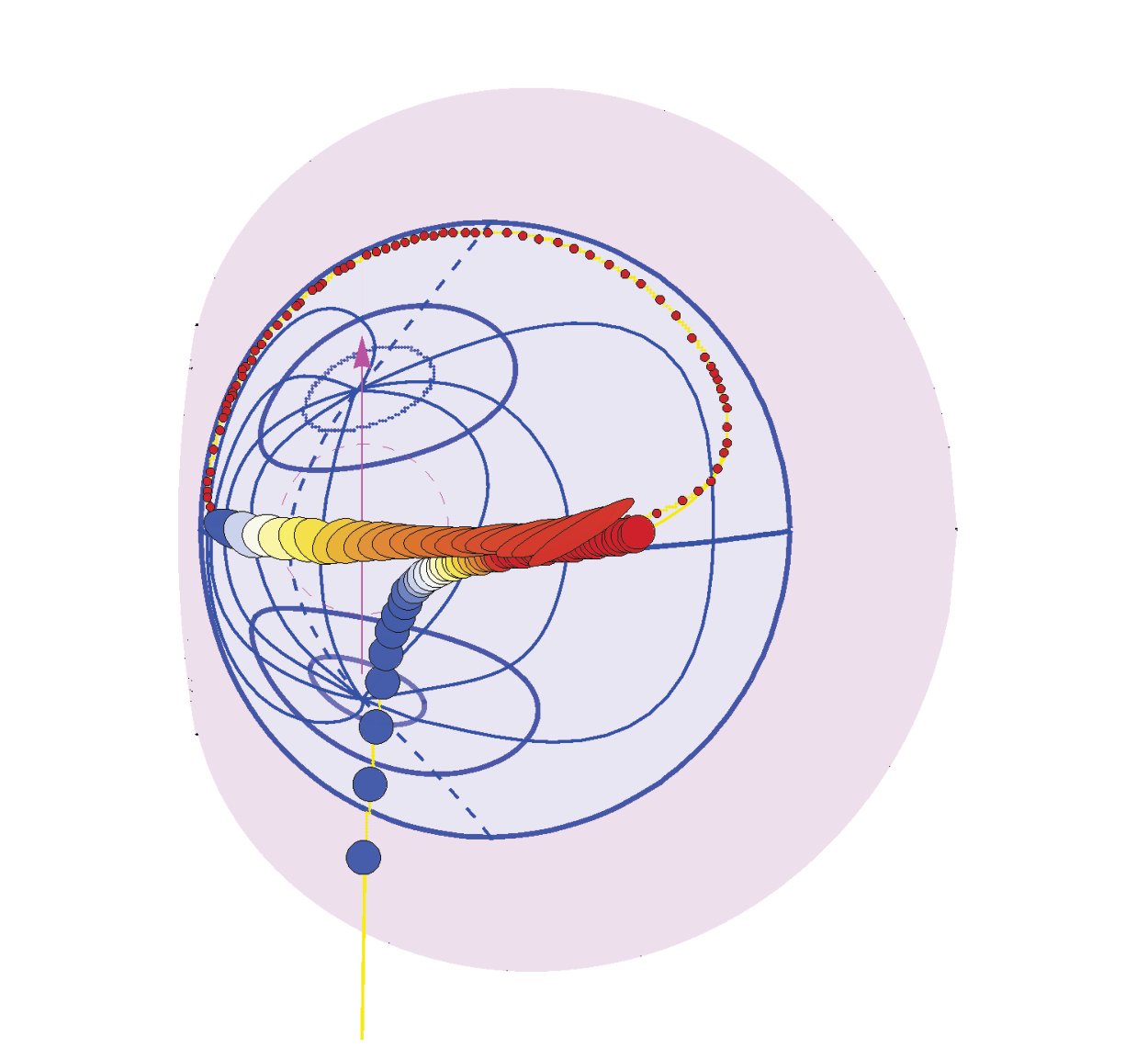,width=11cm}}
\vspace*{8pt}
\caption{The lensed images in discrete times of the spherical compact star with orbital parameters $\gamma=1$, $\lambda=q=0$ plunging at the equatorial plane into rotating black hole with $a=0.998$. A distant observer is situated at $\cos\theta=0.1$. The lensed images are winding in the azimuthal $\phi$-direction with an angular velocity $\Omega_{\rm h}$ from Eq.~(\ref{Omegah}), when plunging star is approaching to the event horizon. The lensed images are gradually fading in time during each winding circle. It is shown the first circle of this winding.  \label{f8}}
\end{figure}

\begin{figure}[pt]
\centerline{\psfig{file=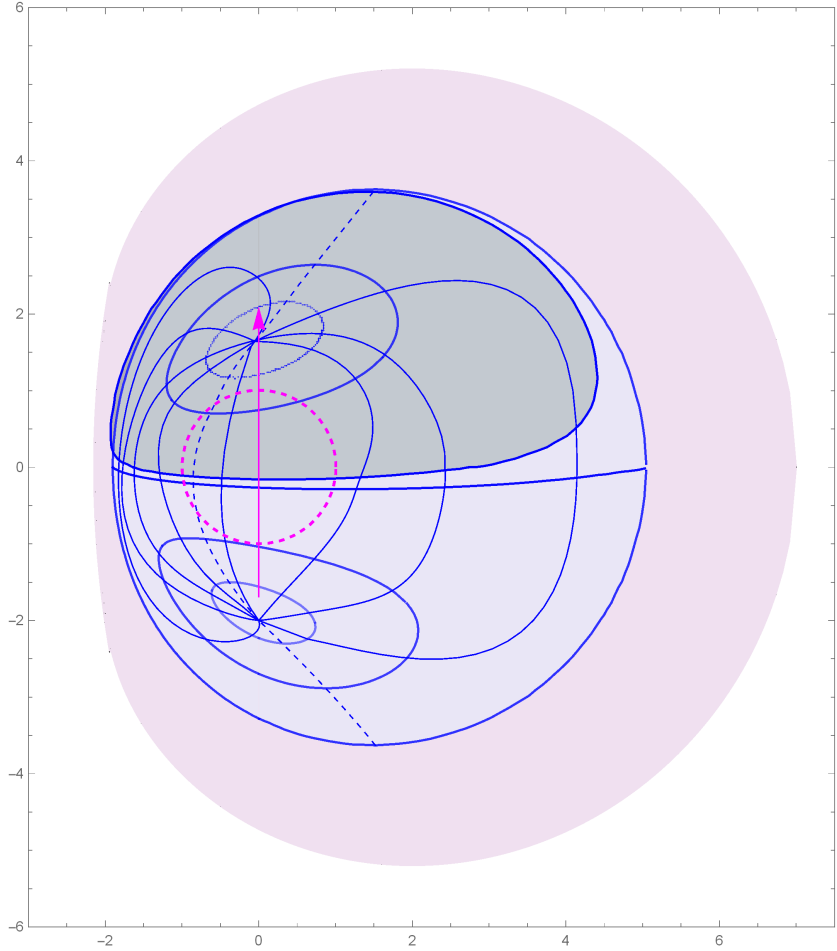,width=11cm}}
\vspace*{8pt}
\caption{Reconstructed event horizon image within a shadow of the rotating black hole with $a=0.998$, viewed on the celestial sphere by a distant observer, situated at $\cos\theta=0.1$. The dark gray region is a northern hemisphere of the event horizon globe. An equatorial parallel is the boundary of this dark region. A dark northern hemisphere of the event horizon is the simplest model of the black hole silhouette in the presence of a thin accretion disk. \label{f10}}
\end{figure}

\begin{figure}[pt]
\centerline{\psfig{file=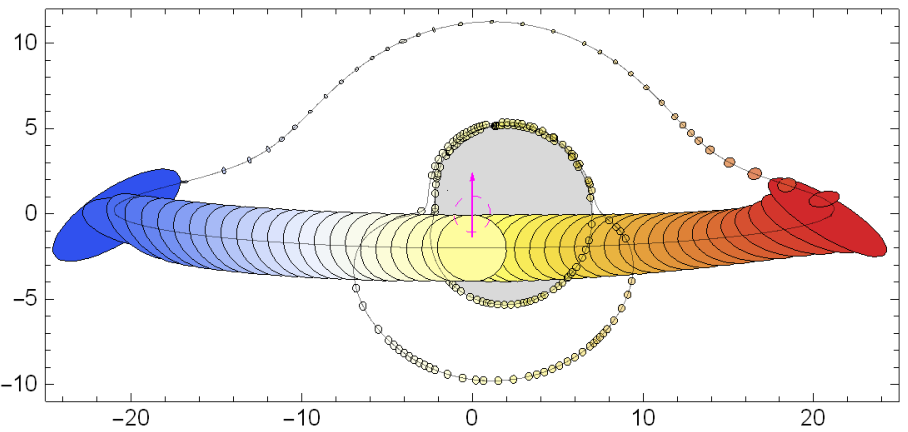,width=11cm}}
\vspace*{8pt}
\caption{Lensed images of a compact luminous probe orbiting the rotating black in the equatorial plane at radius $r=20$ and viewed in discrete time intervals by a distant observer, situated at $\cos\theta=0.1$. Direct probe images and also the first and second light echoes are all placed outside the black hole shadow (filled gray region).  For details see \cite{doknaz18a,doknaz18b}. \label{10}}
\end{figure}

A distant observer will see the gradually faded image of each plunging probe, approaching to the event horizon. Position of the last detected lensed photon from the plunging probe on the celestial sphere will map the corresponding point on the lensed event horizon image. The resulting total lensed image of the event horizon would be the one-to-one projection of the whole event horizon globe on the celestial sphere. This unique property means that distant observers may view at once both the front and back sides of the event horizon. See this unique black hole feature in Figs.~4--9. 

The outer boundary of the lensed event horizon image, viewed by a specific distant observer in the black hole equatorial plane, is defined from solution of the integral equation
\begin{equation}
\int_2^\infty\frac{dr}{\sqrt{V_r}}
=2\int_{\theta_0}^{\pi/2}\frac{d\theta}{\sqrt{V_\theta}}.
\label{a0max}
\end{equation}
where $\theta_0$ is a turning point in the latitudinal $\theta$-direction along the photon trajectory for direct lensed image, defined from equation $V_\theta=0$, where $V_\theta=0$ is from Eq.~(\ref{Vtheta}). According to the Cunnungham--Bardeen classification scheme for multiple lensed images \cite{CunnBardeen72,CunnBardeen73} the photons producing direct lensed images do not intersect the black hole equatorial plane on their way from emitting probe to a distant observer.

In the Schwarzschild case with $a=0$ a turning point on the escaping photon trajectory in $\theta$--direction is at $\theta_0=\arccos(q/\sqrt{q^2+\lambda^2})$ and the value of the right-hand-side integral in (\ref{a0max}) is $\pi//\sqrt{q^2+\lambda^2}$. With this simplification, from numerical solution of integral equation (\ref{a0max}) we find the radius of the event horizon image $r_{\rm eh}=\sqrt{q^2+\lambda^2}=4.457$. The near (light blue) hemisphere of the event horizon globe with a radius $r_{\rm h}=2$ is projected by the lensing photons into the central (light blue) disk with a radius $r_{\rm EW}\simeq2.848$. Respectively the far (dark blue) hemisphere is projected into the hollow (dark blue) disk with an outer radius $r_{\rm eh}\simeq4.457$, which is a radius of the lensed event horizon image in the Schwarzschild case. 

See in Fig.~4 the trajectories of some photons forming the lensed event horizon image of the Schwarzschild black hole and images of some parallels (blue curves) and meridians (black curves) of the event horizon globe. A fictitious outer border of the event horizon in the fictitious Euclidean space is shown by the dashed magenta circle in Fig.~4 and in the subsequent Figs. In Fig.~5 it is shown the lensed event horizon image of the Schwarzschild black hole projected on the celestial sphere. A distant observer views at once both the front and back sides of the event horizon.

In the general Kerr case, when $a\neq0$, the corresponding turning point on the escaping photon trajectory in $\theta$--direction is at
\begin{equation}
\label{thetamax}
\theta_{0}=\arccos\left[\sqrt{\frac{\sqrt{4a^2q^2+(q^2+\lambda^2-a^2)^2}
	-(q^2+\lambda^2-a^2)}{2a^2}}\;\right]\!, 
\end{equation} 
and the event horizon image has a more complicated form with respect to the simple disk. A corresponding numerical solution of integral equation  (\ref{a0max}) for the outer boundary of the event horizon image of the extreme Kerr black hole with $a=1$ is shown graphically in Figs.~6 and 7. In these Figs. a near (light blue) hemisphere of the event horizon globe is projected by the lensing photons into the central (light blue) region. Respectively, the far (dark blue) hemisphere is projected into the hollow (dark blue) region. Images of some parallels (blue curves) and meridians (black curves) of the event horizon globe ar also shown. The dashed curve in Fig.~6 and in the subsequent Figs. is an instant prime meridian. In the general Kerr case, when $a\neq0$, as well as in the Schwarzschild case, the distant observer views at once both the front and back sides of the event horizon. 

The specific feature of test particle trajectories in the general Kerr case, when $a\neq0$, is that the plunging test probes are winding in the azimuthal $\phi$-direction by approaching to the event horizon. This peculiarity considerably complicates the procedure for reconstruction of meridians on the event horizon image. We choose the ad hoc definition: the lensed image of meridians on the rotating event horizon globe are marked by photons, emitted at the same radii along the chosen meridian, $r=r_{\rm h}+\epsilon$ and $\epsilon=const\ll1$, i.\,e., very close to the event horizon at $r=r_{\rm h}$. Once being marked, these lensed image of meridian will be rotating synchronously with the event horizon. The black hole horizon and its lensed image are both synchronously rotating with an angular velocity $\Omega_{\rm h}$ from Eq.~(\ref{Omegah}).

See in Fig.~8 the lensed images of compact star plunging into rotating black hole with $a=0.998$, which are shown in discrete time intervals and viewed by a distant observe, situated at $\cos\theta=0.1$. These lensed images are winding around black hole in the azimuthal $\phi$-direction when plunging star is approaching to the black hole event horizon. In particular, the lensed photons from this winding reveal asymptotically, at $t\to\infty$, the equatorial parallel of the event horizon image.
See details of this numerical animation in Ref.~ \refcite{doknaz18c}.

In Fig.~9 the equatorial parallel of the event horizon image is a boundary of the dark region, which is the reconstruction on celestial sphere of the northern hemisphere of the event horizon globe. Note that a dark northern hemisphere of the event horizon is the simplest model for a black hole silhouette in the presence of a thin accretion disk.

For comparison with the lensed images of the stationary moving probe see in Fig.~10 the numerical calculation of the lensed direct image and the first and second light echoes images of a compact probe (star) stationary orbiting around rotating black hole and viewed by a distant observer \cite{doknaz18a,doknaz18b}. Note that all lensed multiple images (light echoes)) of the stationary moving luminous probe are placed outside the black hole shadow.

\section{Conclusion}

We demonstrate that gravitational lensing of the luminous objects plunging into black hole provides the principle possibility for reconstruction of the black hole event horizon image on the celestial sphere. It is appropriate to call this reconstructed projection of the event horizon on the celestial sphere for a distant observer as  the ``{\sl lensed event horizon image}'', or simply the ``{\sl event horizon image}''. This event horizon image is placed on the celestial sphere within the position of black hole shadow and may be considered as a genuine silhouette of the black hole.

Amazingly, the event horizon image is a gravitationally lensed projection on the celestial sphere of the whole surface of the black hole event horizon globe. In result, the black holes may be viewed at once from both the front and back sides. The similar statements on the properties of the event horizon image for Schwarzschild case were made in \cite{Schwar}. A dark northern hemisphere of the event horizon image is the simplest model for a black hole silhouette in the presence of a thin accretion disk.

Note, that results of this work are suitable only in the case of a slow (nonrelativistic) motion of the observer with respect to the black hole. See Ref. \refcite{KopMash02} for a promising formalism of light propagation in the gravitational field of self-gravitating spinning bodies moving with arbitrary velocities.

\section*{Acknowledgments}

We are grateful to E. O. Babichev, V. A. Berezin, Yu. N. Eroshenko and A. L. Smirnov for stimulating discussions.

This work was supported in part by the Russian Academy of Sciences Program of basic research 12 "Problems of Origin and Evolution of the Universe".

\end{document}